\documentclass[12pt,twoside]{article}
\usepackage{fleqn}
\usepackage{espcrc1}
\usepackage{graphicx}
\usepackage[figuresright]{rotating}
\usepackage{wrapfig}

\title{Effective Weak Chiral Lagrangian from the Instanton Vacuum
\thanks{Talk given by HCK at PANIC `99, Uppsala Sweden, June 10, 1999.}
\\[-4cm]
\vskip-4cm
\begin{flushright}
\normalsize
PNU-NTG-03/99 \\ RUB-TPII-10/99 \\
July 1999
\end{flushright}
\vskip1.5cm
}
\author{Mario Franz\address{
Institute for Theoretical  Physics  II,
Ruhr-University Bochum, D-44780 Bochum, Germany}
\footnote{email:mariof@tp2.ruhr-uni-bochum.de},
Hyun-Chul Kim\address{
Department of Physics, Pusan National University,
609-735 Pusan, Republic of Korea}\thanks{Research supported by 
the Korea Research Foundation made in the program year of 
1998}\footnote{email:hchkim@hyowon.pusan.ac.kr},
and Klaus Goeke$^{\rm a}$
\footnote{email:goeke@tp2.ruhr-uni-bochum.de}
}

\begin{document}

%\bibliographystyle{prsty}
%\twocolumn[
\maketitle
%\widetext
\begin{abstract}
We present in this talk the recent investigation of the 
effective weak chiral Lagrangian within the framework of
the instanton-induced chiral quark model.  The low energy constants 
in leading order $g_8$ and $g_{27}$, and their ratio are determined.
The momentum-dependence of the constituent quark mass turns out to
improve the low energy constants to the great extent, respecting the
$\Delta T=1/2$ enhancement.
\end{abstract}

\medskip
\noindent\rule{2.5in}{.5pt}
\bigskip

{\bf 1.} Understanding the $\Delta T = 1/2$ selection rule, 
best known as the fact that the isospin 
amplitude of the $K\rightarrow \pi\pi$ decay is about 22 times 
larger than the $T=2$ amplitude, remains still elusive
in hadronic physics.  Despite a great deal of effort this 
dominance of the $\Delta T=1/2$ channel over the $\Delta T=3/2$ one 
has not been explained in a satisfactory manner.  While a part of the answer  
comes from perturbative gluons created by the evolution from
a scale of $M_W\simeq 80$ GeV to around 1 GeV, another part of the answer
is supposed to arise from the structure of the light hadrons, whose
description at scales around 1 GeV requires a method of nonperturbative QCD.

In chiral perturbation theory ($\chi$PT)~\cite{Kamboretal,Esposito,EKW}
the $\Delta T=1/2$ enhancement is represented by the ratio of the
low energy constants $g_8$ and $g_{27}$:
$|g_8 +\frac15 g_{27}|_{\rm exp}\simeq 5.1,
\;|g_{27}|_{\rm exp}\simeq 0.16$~\cite{PichRafael}.  
Kambor {\em et al.} showed 
that the octet coupling constant is decreased by $30\%$ with the chiral 
loops considered.  So, the ratio is reduced to about $22$.  
However, the inclusion of the chiral loops ${\cal O}(p^4)$ introduces 
so many low energy constants (LECs) that it is not possible to constrain the
LECs to order ${\cal O}(p^4)$ based on the experimental data to date.      
One need to resort to effective QCD-inspired models in order to proceed
without relying on experimental results.

Recently, M. Franz {\em et al.}~\cite{FranzKimGoeke}
have investigated the effective 
$\Delta S=1,2$ weak chiral Lagrangian to order ${\cal O}(p^4)$ 
within the framework of the chiral quark model 
($\chi$QM), focusing on determining the LECs in the effective weak chiral 
Lagrangian.  However, the 
ratio of the $g_8/g_{27}$ obtained from the $\chi$QM are deviated from the
phenomenological values from $\chi$PT.   

In this talk, we want to present the recent improvement of
the former study~\cite{FranzKimGoeke}, 
considering the momentum dependence of the mass of
the constituent quark which considered as a constant in the former 
work.  The momentum-dependent dynamical quark mass arises from  
the picture of the instanton vacuum which pertains to nonperturbative QCD.  
The instanton vacuum elucidates one of the most important low-energy 
properties of QCD, {\em i.e.} the mechanism of spontaneous breaking of 
chiral symmetry~\cite{DP}, which brings about 
the momentum-dependent quark mass.

{\bf 2.} 
The low-energy QCD partition function in Euclidean space can be written
as  
\begin{equation}
{\cal Z}\;=\; \int {\cal D} \psi {\cal D} \psi^\dagger 
{\cal D} \pi^a
\exp \left[\int d^4 x \psi^{\dagger \alpha}_{f}
\left(i\rlap{/}{\partial} + 
i\sqrt{M(-i\partial)} U^{\gamma_5}
\sqrt{M(-i\partial)}\right)_{fg}
\psi^{\alpha}_{g} \right],
\label{Eq:Dirac}
\end{equation}
The $M(k)$ is the momentum-dependent constituent quark mass expressed 
as follows:
\begin{equation}
M(k) \;=\; M_0 F(k)^2
\end{equation}
The $F(k)$ is normalized to unity at $k=0$.  Thus, $M_0$ is the value of
the $M(k)$ at $k=0$.  The momentum-dependent constituent quark mass
can be regarded as a UV regulator.  The pion decay constant $f_\pi$  
is obtained as follows:
\begin{equation}
f_{\pi}^2 \;=\; 4N_c\int \frac{d^4 k}{(2\pi)^4} 
\frac{M^2(k) - \frac12 M(k) \frac{dM(k)}{dk}k 
+ \frac14 \left(\frac{dM(k)}{dk}\right)^2 k^2}{(k^2 +M^2(k))^2} .
\label{Eq:fpiq}
\end{equation}
The $M(k)$ is contrained by reproducing the experimental value
of $f_\pi=93$ MeV.  The quark and gluon condensates are written,
respectively, by
\begin{equation}
\langle \bar{\psi}\psi\rangle\;=\;
-4 N_c \int \frac{d^4 k}{(2\pi)^4}\frac{M(k)}{k^2 +M^2(k)},\;\;\;
\left\langle\frac{\alpha_s}{4\pi}G^a_{\mu\nu}G^{a\mu\nu}\right\rangle
\;=\; 8 N_c \int \frac{d^4 k}{(2\pi)^4}\frac{M^2(k)}{k^2 +M^2(k)}.
\label{Eq:condensate}
\end{equation}

The effective weak chiral Lagrangian $S^{\Delta S = 1}_{\rm eff}[\pi]$ 
in lowest order of $G_F$ can be obtained as follows: 
\begin{equation}
{\cal L}^{\Delta S = 1}_{\rm eff} \;=\; -\frac{1}{\cal N}
\int {\cal D} \psi {\cal D} 
\psi^\dagger {\cal H}^{\Delta S = 1}_{\rm eff} \exp  
\left[\int d^4 x \psi^\dagger \left(
i\rlap{/}{\partial} + i\sqrt{M(-i\partial)} U^{\gamma_5}
\sqrt{M(-i\partial)} \right)\psi\right].
\label{Eq:part1}
\end{equation}
Here the effective weak quark Hamiltonian 
${\cal H}^{\Delta S = 1}_{\rm eff}$ consists of ten four-quark operators 
among which only seven operators are independent:
\begin{equation}
{\cal H}^{\Delta S = 1}_{\rm eff}
 \;=\; -\frac{G_F}{\sqrt{2}} V_{ud} V^*_{us}
\sum_i c_i (\mu) {\cal Q}_i (\mu) + {\rm h.c.} .
\end{equation}
The expression for the four-quark operators ${\cal Q}_i$ and Wilson
coefficients can be found elsewhere~\cite{Burasetal}.

Integrating over the quark fields and using the derivative expansion,
the effective weak chiral Lagrangian in leading order is then obtained as
follows:
\begin{eqnarray}
{\cal L}^{\Delta S = 1,{\cal O} (p^2)}_{\rm eff}&=& 
-\frac{G_F}{\sqrt{2}} V_{ud} V^*_{us} f^4_{\pi} 
\left[g_{8} \left\langle\lambda_{23} L_\mu L^\mu \right\rangle
\right.\nonumber \\ && \;+\; \left.g_{27} \left(\frac23
\left\langle \lambda_{12} L_\mu\right\rangle
\left\langle \lambda_{31} L^\mu \right\rangle  + 
\left\langle \lambda_{32} L_\mu\right\rangle
\left\langle \lambda_{11} L^\mu \right\rangle \right) \right]
\;+\; {\rm h.c.}  
\end{eqnarray}
The coupling constants $g_{8}$ and $g_{27}$ 
can be extracted from the $K\rightarrow \pi\pi$ decay rate 
and the $\Delta T = 1/2$ enhancement is reflected in these constants.
In our model those constants are expressed in terms of the Wilson coefficients 
and
the dynamical coefficients ${\cal K}$, ${\cal M}$, ${\cal P}$, and
${\cal R}$~\cite{FranzKimGoeke2} which are functions of the momentum-dependent
quark mass:
\begin{eqnarray}
g^{(1/2)}_{8} &=& \frac{16 N_c^{2}{\cal K}^2}{f^4_{\pi}}
\left(-\frac25 c_1 + \frac35 c_2 + c_4 -\frac35 c_9 + \frac25 c_{10}
\right) +\frac{64 N_{c}^{2} {\cal M} ({\cal P} + {\cal R})}{f_{\pi}^{4}}c_6,
\nonumber \\ 
g^{(3/2)}_{27} &=& 
\frac{16 N_c^{2} {\cal K}^2}{f^4_{\pi}}
\left(\frac35 c_1 + \frac35 c_2 + \frac{9}{10} c_9 
+\frac{9}{10} c_{10}\right).
\label{Eq:ratiol}
\end{eqnarray}

{\bf 3.} In this talk, we will show the results with the original
$M(k)$~\cite{DP} employed, which turns out to be the best for
the present calculation.  For the other types of the $M(k)$, we
refer to the corresponding work~\cite{FranzKimGoeke2}.  
In Figures 1 and 2 the quark and gluon condensates are drawn, 
respectively.        
\begin{figure}[t]
\begin{minipage}[t]{75mm}
\includegraphics[width=70mm,height=50mm]{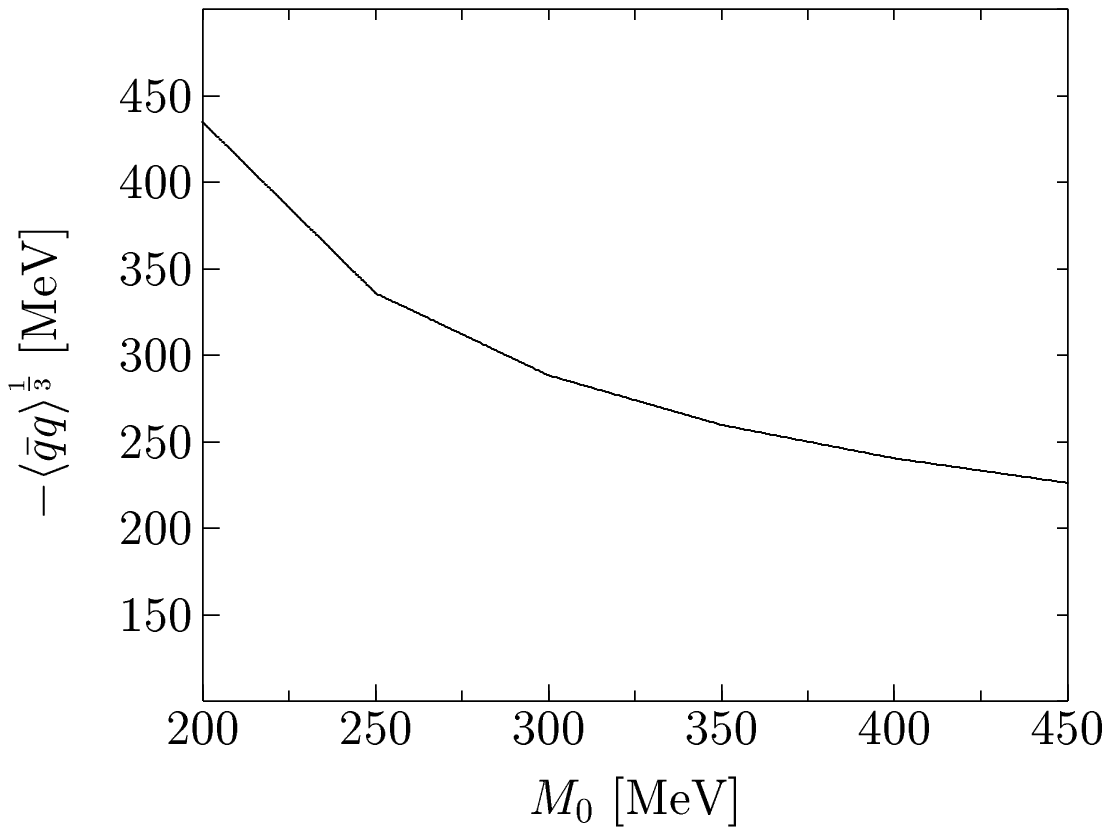}
%\framebox[79mm]{\rule[-26mm]{0mm}{52mm}}
\vspace*{-.3in}
\caption{The quark condensate as a function of $M_0$.}
\label{fig1} 
\end{minipage}
\hspace{\fill}
\begin{minipage}[t]{75mm}
%\framebox[74mm]{\rule[-26mm]{0mm}{52mm}}
\includegraphics[width=70mm,height=50mm]{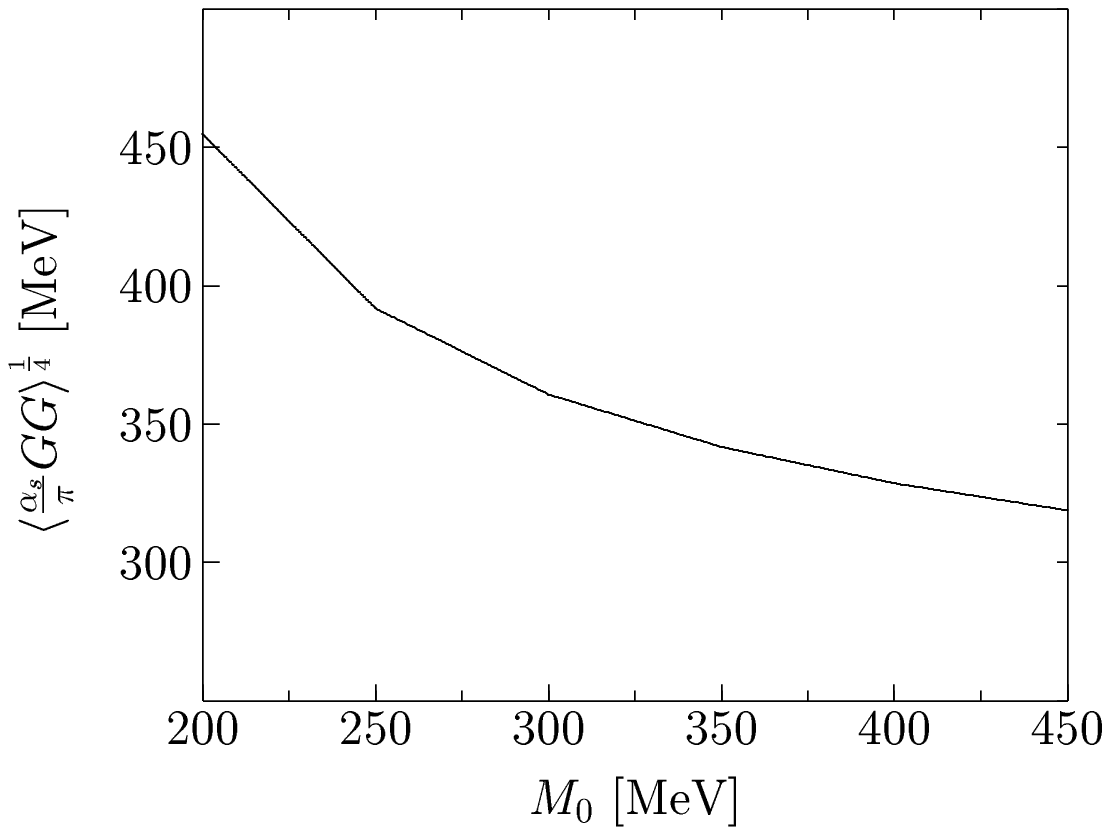}
\vspace*{-.3in}
\caption{The gluon condensate as a function of $M_0$.}  
\label{fig2}
\end{minipage}
\end{figure}
Both of the quark and gluon condensates increase as the $M_0$ decreases.
In particular, they increase relatively faster in smaller values of the $M_0$.

In Figure 3, we draw the low energy constants $g_8$ and $g_{27}$,
respectively.  While the $g_8$ changes drastically as the $M_0$
does, the $g_{27}$ midly decreases as the $M_0$ increases. 
It indicates that the $\Delta T=1/2$ enhancement is 
better explained as we choose smaller values of the $M_0$.

Figure 4 shows the ratio $g_{8}/g_{27}$ of the LECS in leading order,
compared with the former result in the shaded band for which the constant
quark mass ($F^2(k)=1$) is used and the cut-off parameter 
$\Lambda$ is introduced to tame the divergence. 
Note that the quark condensate appears directly in the expression for
the LECs in the former work.  
On the other hand, the quark condensate does not enter in the present
result as seen in Eq.(\ref{Eq:ratiol}).  The coefficient ${\cal M}$ 
in Eq.(\ref{Eq:ratiol}) becomes the quark condensate only 
when we turn off the momentum-dependence of the quark mass, 
which turns out to be identical to the former result.    
The improvement achieved by the momentum-dependent
quark mass is prominent, compared to the former work~\cite{FranzKimGoeke}.
In particular, the improvement is remarkable at smaller values of the $M_0$, 
as we already noticed in Fig. 3.    
\begin{figure}[t]
\begin{minipage}[t]{75mm}
%\framebox[74mm]{\rule[-26mm]{0mm}{52mm}}
\includegraphics[width=70mm,height=50mm]{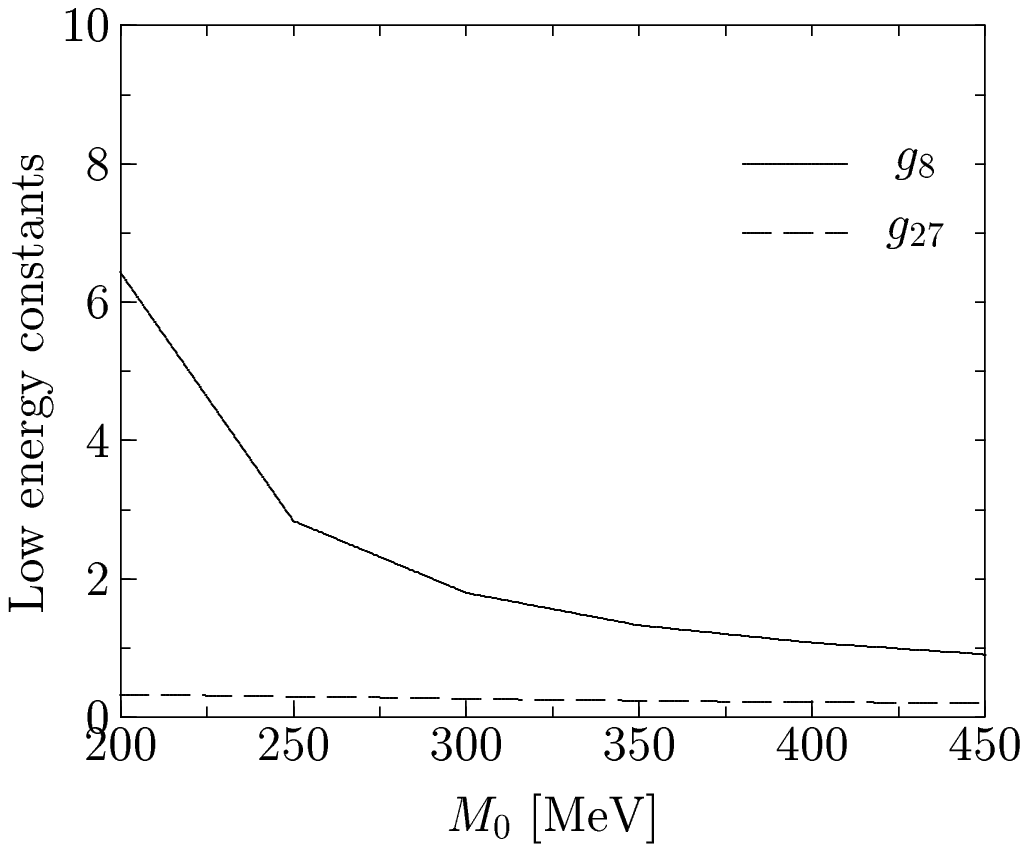}
%\vspace*{-.3in}
\caption{The low energy constants as a function of $M_0$.  The solid curve
draws the $g_8$, while the dashed one represents the $g_{27}$.}
\label{fig3}
\end{minipage}
\hspace{\fill}
\begin{minipage}[t]{75mm}
%\framebox[74mm]{\rule[-26mm]{0mm}{52mm}}
\includegraphics[width=70mm,height=50mm]{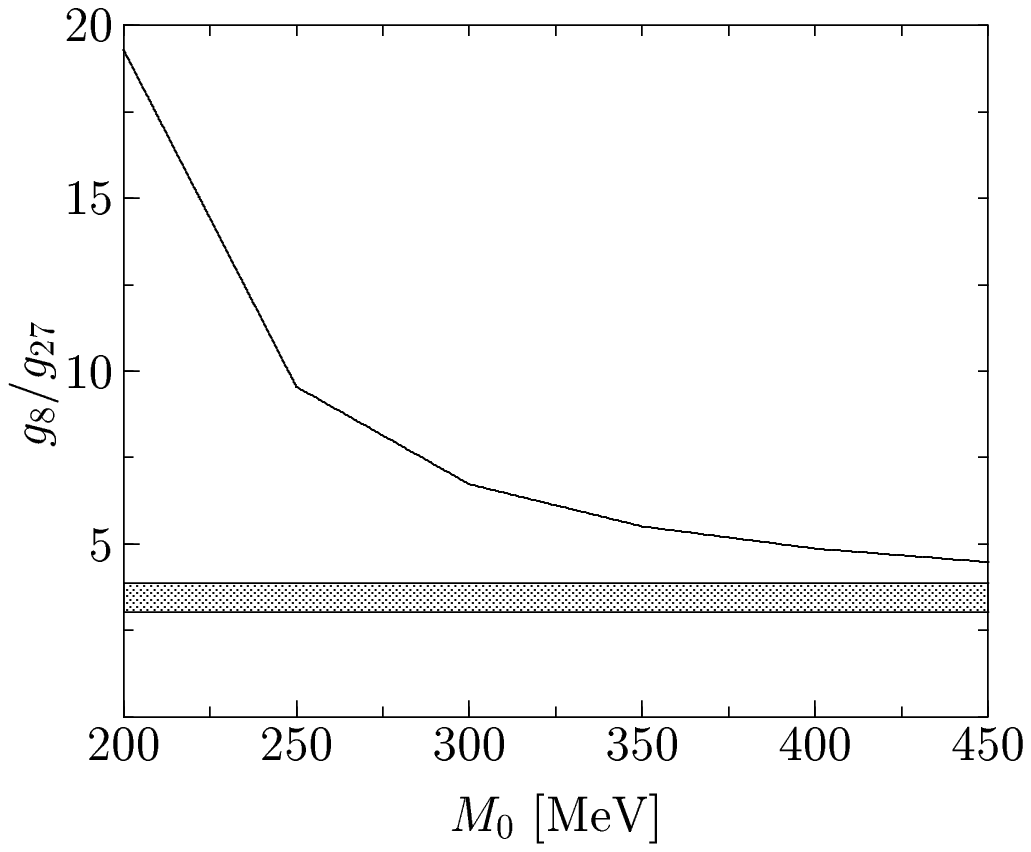}
%\vspace*{-.3in}
\caption{The ratio $g_{8}/g_{27}$ as a function of $M_0$.  The solid curve
draws the present result,
while the shaded band designates the former 
one~\protect{\cite{FranzKimGoeke}} with the uncertainty from  
the range of input parameters.}
\label{fig4}
\end{minipage}
\end{figure}   
For example, with the value of $M_0=200$ MeV, 
we obtain the ratio which is almost 
the same as the empirical result, allowing the quark condensate to
be larger than usual.  However, bearing in mind that there might be other 
sources of the $\Delta T=1/2$ enhancement such as the matching between the 
scale of the physical decay processes and that of the 
weak Hamiltonian~\cite{Bijnens}, or $1/N_c$ corrections ({\em e.g.} 
meson loops), we do not want to insist too much strongly smaller values 
of the $M_0$ so as to reproduce the empirical result of the ratio.    

The conclusion we draw from the present work is that the momentum-dependence 
of the constituent quark mass plays an essential role in improving the 
low energy constants to the great extent, respecting the $\Delta T =1/2$
enhancement.

\end{document}